\documentclass[%
 reprint,
 amsmath,amssymb,
 aps,
]{revtex4-1}

\usepackage{graphicx}% Include figure files
\usepackage{dcolumn}% Align table columns on decimal point
\usepackage{bm}% bold math
\newtheorem{theorem}{Theorem}
\newtheorem{proposition}{Proposition}
\newtheorem{lemma}{Lemma}
\newtheorem{corollary}{Corollary}
\newtheorem{remark}{Remark}

\def\ba{\begin{array}}
\def\ea{\end{array}}
\def\be{\begin{equation}}
\def\ee{\end{equation}}

\def\ds{\displaystyle}

\def\v{{\bf v}}

\def\0{{\bf 0}}
\def\1{{\bf 1}}
\def\2{{\bf 2}}
\def\3{{\bf 3}}
\def\4{{\bf 4}}
\def\5{{\bf 5}}
\def\6{{\bf 6}}
\def\7{{\bf 7}}
\def\8{{\bf 8}}
\def\9{{\bf 9}}

\usepackage{graphicx}\usepackage{dcolumn}\usepackage{bm}\usepackage{amsmath}\usepackage{amscd}\usepackage{amsfonts}\usepackage{extarrows}
\usepackage[colorlinks,
            linkcolor=red,
            anchorcolor=red,
            citecolor=red,
            ]{hyperref}
\usepackage{booktabs}

\def\bt{\begin{theorem}}
\def\et{\end{theorem}}
\def\bp{\begin{proposition}}
\def\ep{\end{proposition}}
\def\bc{\begin{corollary}}
\def\ec{\end{corollary}}
\def\bo{\begin{proof}}
\def\eo{\end{proof}}
\def\bx{\begin{example}}
\def\ex{\end{example}}
\def\br{\begin{remark}}
\def\er{\end{remark}}
\def\bl{\begin{lemma}}
\def\el{\end{lemma}}

\begin{document}

\preprint{APS/123-QED}

\title{From linear combination of quantum states to Grover's searching algorithm}% Force line breaks with \\

\author{Changpeng Shao}
\email{cpshao@amss.ac.cn}
\affiliation{Academy of Mathematics and Systems Science, Chinese Academy of Sciences, \\ Beijing 100190, China}


%\affiliation{
%\affiliation{
%\affiliation{%


\date{\today}% It is always \today, today,

\begin{abstract}
Linear combination of unitaries  (LCU for short) is one of the most important techniques in designing quantum algorithms. In this paper, we propose a new quantum algorithm in three different forms to achieve LCU. Different from previous algorithms \cite{childs-linear-system,clader,long11}, the complexity now only depends on the number of the unitaries and the precision. So it will play more important role in the design of quantum algorithms when the number of unitaries is small, such as quantum iteration algorithms.
%Since in iteration algorithms, $m$ often refers to the iteration steps, which is small if the iteration algorithms are efficient.
Moreover, as an application of the new LCU, three new quantum algorithms to the searching problem will be proposed, which will provide us new insights into Grover's searching algorithm. We also show that the problem of LCU is closely related to the problem of if we can efficiently implement $U^t$ for $0<t<1$ when $U$ is an efficiently implemented unitary operator? This problem is not hard to solve. However, it becomes inefficient when it contains a strict requirement on precision, such as in Grover's algorithm. Finally, as an application of the new LCU technique, we will show that the quantum state of any real classical vector can be prepared efficiently in quantum computer. So this solves the ``input problem" in quantum computer efficiently.
%\begin{description}
%\end{description}
\end{abstract}

\pacs{Valid PACS appear here}% PACS, the Physics and Astronomy
\maketitle

%\tableofcontents

\section{Introduction}
\label{intro}

As one of the most important quantum algorithms, Grover's algorithm \cite{grover96} to the unstructured data searching problem achieves quadratic speedup than any classical algorithm. So any problem relates to searching can be improved directly by Grover's algorithm in quantum computer. More importantly, Grover's algorithm can achieve speedup (sometimes not just quadratic speedup) to many NP complete problems, such as 3-SAT \cite{ambainis}, existence of Hamiltonian cycle \cite{nielsen}, quadratic Boolean equations solving \cite{faugere}, etc. Unfortunately, it has been shown that Grover's algorithm is optimal \cite{bennett,boyer}, so there does not exist more efficient quantum algorithms to solve the searching problem.

From the viewpoint of designing new quantum algorithms, we prefer to focus more on the new techniques, directly or indirectly, derived from Grover's algorithm. These include quantum amplitude amplification technique \cite{brassard-AA}, quantum counting \cite{brassard-counting}, swap test \cite{buhrman}, quantum walk \cite{szegedy,watrous} and so on.  For instance, quantum amplitude amplification almost used in all quantum algorithms to improve the success probability with a quadratic speedup. Swap test achieves an exponential speedup to estimate the inner product of two quantum states. As a generalization of the classical random walk, quantum walk currently becomes one important technique and model to study quantum algorithms. Different from other quantum algorithms (such as HHL algorithm \cite{harrow} and its applications \cite{lloyd,rebentros,rebentros-newton,rebentros-svd,wiebe}) resulting from Shor's algorithm or quantum phase estimation algorithm, the quantum algorithms (such as \cite{ambainis07,buhrmanspalek,magniez}) obtained from quantum walk often contains no restrictions.

As shown in \cite{long06,long11}, Grover's algorithm can be viewed as a simple application of linear combination of unitaries (LCU) in duality quantum computer. The problem of LCU can be stated as follows: given $m$ numbers $\alpha_j$ and $m$ efficiently prepared quantum states $|x_j\rangle$, then how to prepare $|y\rangle$ proportional to $y=\sum_j\alpha_j|x_j\rangle$? Here $|x_j\rangle$ can be some unknown quantum states obtained from some quantum algorithms. This is a very fundamental problem about the operations among quantum states. The quantum algorithms to study this problem have been discovered many years ago, most of them are based on control operations. LCU technique was first proposed by Long \cite{long06} as a basic operation of duality quantum computer. Later works include \cite{clader} which can be viewed as an inspiration of HHL algorithm; \cite{childs} which is a special case of \cite{long06}, but contains plenty of applications in Hamiltonian simulation \cite{berry,childswiebe}.
Currently, LCU contains many applications in the design of quantum algorithms, for example, see \cite{berry,childswiebe,childs-linear-system,harrow,kerenidis,rebentros}. It is also believed that LCU will become more important and prevalent in the design of new quantum algorithms in the future.

In this work, we focus more on the real linear combination of real quantum states, in the hope of finding other useful techniques to design new quantum algorithms. We will give three new ideas to solve the LCU problem in the real field. Similar to Grover's algorithm, the new quantum algorithms proposed here also contain clear geometric interpolations. The requirement of real field is also because of the clear geometric interpolations.  And studying real quantum states are already enough to solve many problems in numerical analysis. Different from the already discovered quantum algorithms \cite{childs-linear-system,long11}, whose complexity is $O(\sum_j|\alpha_j|/\|y\|)$, the complexity now is independent of $\alpha_j$ and the norm of the $y$. More precisely, the complexity of our new proposed quantum algorithm to achieve LCU is $O(m^{\log (m/\epsilon)})$, where $\epsilon$ is the precision. So if $m$ is small, then our algorithm seems much better. The new LCU also perform better than the old LCU when $|\alpha_j|$ is very large.
Note that in iteration algorithms, $m$ often refers to the iteration step, which is small if the iteration methods are efficient. So the new LCU technique is quite important in the design of quantum iteration algorithms, for instance see \cite{kerenidis,rebentros,shao}.  Especially in the work of quantum Arnoldi method \cite{shao}, the new LCU method performs much better than the old LCU methods.

Another application of the new LCU technique we will consider in the work is the ``input problem" in quantum computer \cite{biamonte}.
The input problem can be stated as the transformation from classical data $x=(x_0,\ldots,x_{n-1})$ into quantum data $|x\rangle = \frac{1}{\|x\|} \sum_{j=0}^{n-1} x_j|j\rangle$.
If forms the first step of many quantum algorithms, such as HHL algorithm \cite{harrow} and its applications \cite{lloyd,rebentros,rebentros-newton,rebentros-svd,wiebe}. Although we already have quantum RAM to solve the input problem efficiently, it is not practical. The input problem can be solved efficiently in certain special cases, such as when the entries of the classical vector are almost in the same size \cite{aaronson,clader} or when $\sum_{j=i_1}^{i_2}|x_j|^2$ can efficiently calculated for all $i_1<i_2$ \cite{grover-state-preparation}.
In this paper, we will show that the input problem can be solved efficiently in polynomial time in the real case by the new LCU technique. The result obtained from the old LCU technique also achieves exponential speedup than the old quantum algorithms to solve the input problem in the general case.

In the special case of LCU when there are only two real quantum states $|a\rangle,|b\rangle$ and we want to prepare $|c\rangle$ proportional to $|a\rangle+|b\rangle$, the three new algorithms to this problem give three new searching algorithms with the same complexity as Grover's algorithm. Actually, this has been discovered very easier in Long's work \cite{long06}, but here we show three new ones.
Note that in \cite[section 6.2]{nielsen}, Nielsen and Chuang give one explanation about Grover's algorithm from the point of Hamiltonian simulation. The three new quantum algorithms to the searching problem can also be viewed as three new explanations about Grover's algorithm from different perspectives.
%Note that in the original Grover's algorithm, if we perform too many rotations, the amplitude of the marked items will decrease. Although there exists some methods \cite{kaye} to overcome this disadvantage, it is not a problem in our three new searching algorithms. This can be viewed as an advantage of searching algorithm based on LCU.

We also find that the above special case of LCU is closely related to the problem of implementing $U^t$ by giving the information of the unitary operator $U$. When $t$ is an integer, then $U^t$ can be implemented in time $O(tC_U)$, where $O(C_U)$ is the implementing complexity of $U$. When $t$ is not an integer, we just need to focus on the case $0<t<1$.
Actually based on quantum phase estimation, this problem is still not hard to solve. However, it may inefficient when the problem contains a strict requirement on the precision, such as the searching problem. As an application of the optimality of Grover's algorithm, we will show that this problem can be solved optimally in time $O(C_U/\epsilon)$.

The structure of this paper is as follows:
In section \ref{Grover's searching algorithm: brief review}, we briefly review the famous Grover's algorithm.
In section \ref{Four methods to achieve the linear combination of two quantum states}, we study the problem of achieving linear combination of two real quantum states with four different methods.
In section \ref{The Grover searching problem}, we show the application of the four methods in the searching problem.
In section \ref{Linear combination of multiple quantum states}, we consider the general case of preparing the quantum state of the linear combination of multiple quantum states in the real field.
Finally, in section \ref{Application of LCU in quantum state preparation}, we show that quantum state preparation problem can be solved efficiently by the new LCU technique.

{\em Notations.} In this paper, $\|\cdot\|$ always refers to the 2-norm of vectors and $i$ refers to the imaginary unit of complex field.

\section{Grover's searching algorithm: brief review}
\label{Grover's searching algorithm: brief review}

Grover's algorithm provides a polynomial speedup over any classical algorithm for a large class of problems. In this section, we briefly review the basic idea  of Grover's algorithm.
Let $f$ be a map from $\mathbb{Z}_N$ to $\mathbb{Z}_2$. Assume that there exists a subset $\mathcal{M}$ (unknown) of $\mathbb{Z}_N$, such that $f(x)=0$ if and only if $x\in \mathcal{M}$. The elements in $\mathcal{M}$ are called marked items. Then the searching problem aims at finding one marked item if $\mathcal{M}$ is not empty. Classical algorithm solves this problem requires about $O(N/M)$ evaluations of $f$, where $M=\#(\mathcal{M})$. However, Grover's algorithm can find one marked item with a high probability using just $O(\sqrt{N/M})$  evaluations of $f$.

The basic structure of Grover's algorithm is very simple. Define the oracle $\mathcal{O}_f$ as
\[
\mathcal{O}_f|x\rangle = (-1)^{f(x)} |x\rangle.
\]
Set
$$|\phi\rangle = \frac{1}{\sqrt{N}} \sum_{x=0}^{N-1} |x\rangle$$
as the superposition of all base states, which is achieved by Hadamard operation. Then Grover's algorithm is applying the rotation $G=(2|\phi\rangle\langle\phi|-I)\mathcal{O}_f$ about $\sqrt{N/M}$ times on $|\phi\rangle$. Finally we will have a high probability close to 1 to find one marked item by measuring. Grover's algorithm contains a clear geometric explanation. The rotation $G$ gradually rotates the superposition state $|\phi\rangle$ close to the quantum states of marked items, which amplifies the amplitude of marked items gradually. Later, this idea was generalized as amplitude amplification technique \cite{brassard-AA} to amplify the amplitude of the target states we want.

To some sense, the above algorithm needs to know the number of marked items, since if we apply the rotation $G$ too many times over a line, the success probability will decrease. The number of applying $G$ depends on the number of marked items.
However, if we do not know this information in advance, there also exist some technique to do the searching by gradually increasing the precision in the quantum phase estimation algorithm \cite{kaye}. The complexity will not change. On the other hand, if we know the number of marked items exactly, the searching algorithm can be modified into exact \cite{long01}.

\section{Four methods to achieve the linear combination of two quantum states}
\label{Four methods to achieve the linear combination of two quantum states}

In this section, we study the LCU problem in its simplest case:
Given two real quantum states $|a\rangle,|b\rangle$, which can be prepared in efficiently time $O(T_{\textmd{in}})$, then how to prepare the quantum state $|c\rangle$ proportional to $|a\rangle+|b\rangle$ and what is the corresponding complexity? In the following, we will propose three new methods (method 2,3,4) to this problem. Before that, we first review the already discovered method, that is method 1 below.

{\bf Method 1.} The first method is based on Hadamard transformation, just like the phase estimation method considered by Kitaev (see \cite{kitaev}, \cite[problem 5.3]{nielsen}), as follows:
\[\ba{lll} \vspace{.2cm}
|0\rangle|+\rangle &\mapsto& \ds \frac{1}{\sqrt{2}} (|0\rangle|a\rangle + |1\rangle |b\rangle) \\
&\mapsto& \ds \frac{1}{2} |0\rangle ( |a\rangle + |b\rangle)+\frac{1}{2} |0\rangle ( |a\rangle - |b\rangle).
\ea \]
Perform measurements, if we get $|0\rangle$, then the post measurement state is $|c\rangle$. The complexity is $O(T_{\textmd{in}}/\||a\rangle + |b\rangle\|)$. This complexity is related to the 2-norm of $|a\rangle + |b\rangle$, which is not good to the algorithm when it is small. In the following, we consider the simplest LCU problem from the viewpoint of geometry.

\begin{figure}[h]
\centering
\includegraphics[width=6cm]{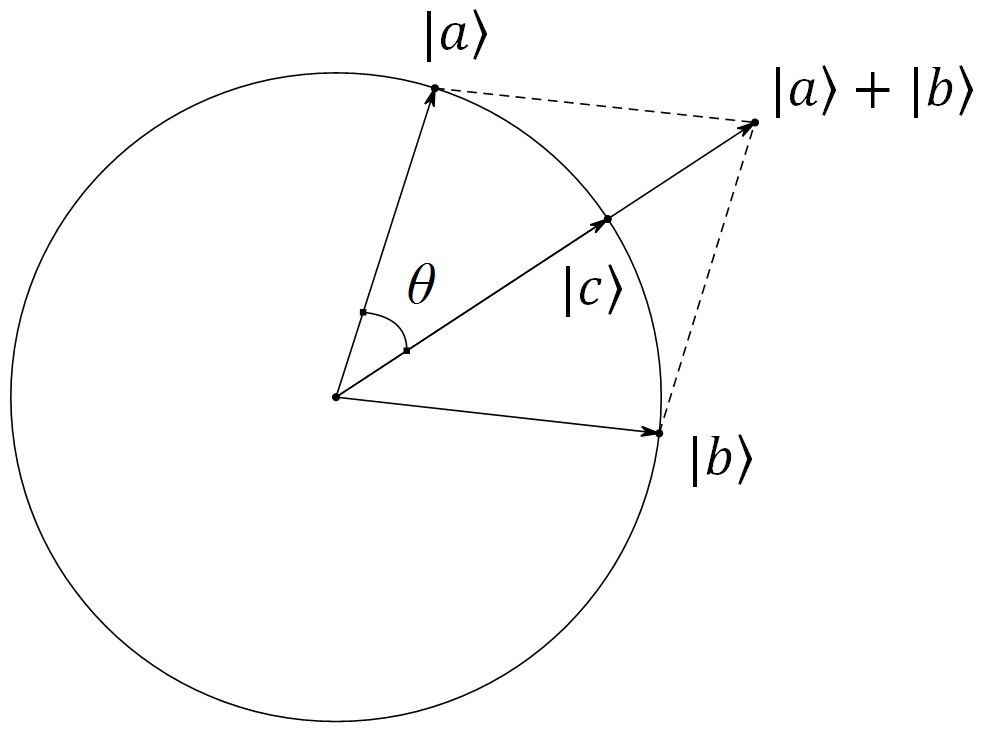}
\caption{Linear combination of two quantum states}
\label{lcs}
\end{figure}

{\bf Method 2.} Denote the angle between $|a\rangle$ and $|c\rangle$ as $\theta$ (see figure \ref{lcs}), then the angle between $|a\rangle$ and $|b\rangle$ equals $2\theta$. For any angle $\phi$, denote the clockwise rotation with angle $\phi$ in the plane spanned by $|a\rangle,|b\rangle$ by $R_\phi$. Then
\[
R_{4\theta}=(I-2|b\rangle\langle b|)(I-2|a\rangle\langle a|).
\]
Therefore, we have $|c\rangle = R_{\theta} |a\rangle = R_{4\theta}^{1/4}|a\rangle$. Now here comes another problem: how to implement $R_{4\theta}^{1/4}$? Generalize this, we actually need to solve the following problem:
Let $U$ be an unitary operator that can be efficiently implemented in time $O(C_U)$, then how to implement $U^t$ for any $0<t<1$ and what is the corresponding implementation complexity? Actually, this problem is not so hard to solve. In the following, we give three methods toward this problem. Two of them can only solve the special case when $U=R^{4\theta}$, another one solves the general case.

From the theory of compact Lie group, we know that there is a unique Hermitian matrix $A$ such that $U=e^{-i A}$. So $U^t=e^{-iAt}$. If we can obtain $A$ efficiently, then $U^t$ will be efficiently implemented due to Hamiltonian simulation, since we have assumed that $U=e^{-i A}$ is efficiently implemented.

Assume that the eigenvalue decomposition of $U=VDV^\dag$, where $D$ is diagonal and $V$ is unitary. Then $A=i\log U=iV (\log D) V^\dag$. Now we consider the case that $U=R_{4\theta}$. The eigenvalues of $R_{4\theta}$ have the form
\[
e^{i\phi},e^{-i\phi},\underbrace{1,\ldots,1}_{n-2},
\]
and the corresponding eigenvectors are $|v_0\rangle,|v_1\rangle,|v_2\rangle,\ldots,$ $|v_{n-1}\rangle$, where $|v_0\rangle=\alpha_0|a\rangle+\beta_0|b\rangle,|v_1\rangle=\alpha_1|a\rangle+\beta_1|b\rangle$ and $|v_2\rangle,\ldots,|v_{n-1}\rangle\in \textmd{span}\{|a\rangle,|b\rangle\}^\bot$.
Then
\[
D=\textmd{diag}\{e^{i\phi},e^{-i\phi},1,\ldots,1\}~\textmd{and}~
V=\sum_{j=0}^{n-1} |v_j\rangle\langle j|.
\]
So we have $\log D=\textmd{diag}\{i\phi,-i\phi,0,\ldots,0\}$ and
\be \label{eq of A}
A=iV (\log D) V^\dag=-\phi(|v_0\rangle\langle v_0|-|v_1\rangle\langle v_1|).
\ee

So $A$ is totally determined by $\phi$ and $\alpha_0,\beta_0,\alpha_1,\beta_1$.
Now we focus on the calculation of these parameters. It is not difficult to show that
\[\ba{lll} \vspace{.2cm}
R_{4\theta}|a\rangle &=& -|a\rangle+2\langle a|b\rangle |b\rangle,  \\
R_{4\theta}|b\rangle &=& -2\langle a|b\rangle |a\rangle+(4\langle a|b\rangle^2-1) |b\rangle.
\ea\]
Note that $\langle a|b\rangle=\cos 2\theta$, since they are real. So the matrix representation of $R_{4\theta}$ in the plane $\textmd{span}\{|a\rangle,|b\rangle\}$ is
\[
R_{4\theta}=\left[
              \begin{array}{cc} \vspace{.3cm}
                -1            &~~~ -2\cos 2\theta \\
                2\cos 2\theta &~~~ 4(\cos 2\theta)^2-1 \\
              \end{array}
            \right].
\]

The two eigenvalues are $e^{\pm 4\theta}$, which implies $\phi=4\theta$. And the corresponding eigenvectors are
\[\ba{lll} \vspace{.2cm}
|v_0\rangle &=& \ds\frac{1}{\sqrt{2}\sin 2\theta} \Big(|a\rangle -e^{i2\theta}|b\rangle\Big), \\
|v_1\rangle &=& \ds\frac{1}{\sqrt{2}\sin 2\theta} \Big(|a\rangle -e^{-i2\theta}|b\rangle\Big).
\ea\]
Finally, a simple calculation on (\ref{eq of A}) with the above results shows that
\be
A=-\frac{4\theta i}{\sin 2\theta} \Big(|a\rangle\langle b|-|b\rangle\langle a|\Big).
\ee
Therefore, $A$ is totally determined by $\theta$, since $|a\rangle,|b\rangle$ are given.
Apply swap test on $|a\rangle,|b\rangle$, the angle $\theta$ can determined in time $O(T_{\textmd{in}}/\epsilon)$ to precision $\epsilon$. Finally, based on the Hamiltonian simulation \cite{nielsen}, $U^t$ can be implemented in time $O(T_{\textmd{in}}/\epsilon)$. So $|c\rangle$ can be prepared in time $O(T_{\textmd{in}}/\epsilon)$ too.

{\bf Method 3.} The third method actually focus on the general problem of implementing $U^t$ based on quantum phase estimation. Denote the eigenvalue decomposition of $U=\sum_{j=0}^{n-1} e^{i\theta_j}|v_j\rangle\langle v_j|$. Then for any quantum state $|d\rangle=\sum_{j=0}^{n-1} \beta_j|v_j\rangle$, we have $U^t|d\rangle=\sum_{j=0}^{n-1} \beta_je^{i\theta_jt}|v_j\rangle$. By quantum phase estimation algorithm, we can get the following precess
\[
\sum_{j=0}^{n-1} \beta_j|v_j\rangle|0\rangle
\mapsto\sum_{j=0}^{n-1} \beta_j|v_j\rangle|\tilde{\theta}_j\rangle
\mapsto\sum_{j=0}^{n-1} e^{i\tilde{\theta}_jt}\beta_j|v_j\rangle|0\rangle,
\]
where $|\tilde{\theta}_j-\theta_j|\leq \epsilon$. The complexity of the above procedure is $O(C_U/\epsilon)$. Moreover, a simple estimation shows that, the error between $\sum_{j=0}^{n-1} e^{i\tilde{\theta}_jt}\beta_j|v_j\rangle$ and $\sum_{j=0}^{n-1} e^{i\theta_jt}\beta_j|v_j\rangle$ is bounded by $O(\epsilon t)=O(\epsilon)$ due to $0<t<1$. Finally, $U^t$ can be implemented in time $O(C_U/\epsilon)$. As for the original problem with $U=R_{4\theta}$, we know that $O(C_U)=O(T_{\textmd{in}})$ and so $|c\rangle$ can be prepared in time $O(T_{\textmd{in}}/\epsilon)$ to precision $\epsilon$.

{\bf Method 4.} The last method that can be used to solve the original problem is as follows. Since it is not so straightforward to implement $R_{4\theta}^{1/4}$, but it is easy to implement $R_{4\theta}^k$ for any positive integer $k$ in time $O(k T_{\textmd{in}})$. Notice that $R_\theta=R_{\theta+2\pi}$, so there is a number $l>1$ such that $R_{4\theta}^{1/4}=R_{4\theta}^l$. Whence we get such a $l$, we can just choose $k$ as the integral part of $l$, that is $k=\lfloor l\rfloor$. As we know, to make the complexity minimal, $l$ equals the smallest number such that $4l\theta \mod 2\pi$ is $\theta$.
Note that this method is quite similar to Grover's searching algorithm, that is we perform a lot of rotations to close to a desired rotation.

{\em Remarks.}
(1). Note that method 1 applies amplitude amplification technique, which is based on Grover's algorithm. Although,
method 2 and 4 need the data of angle $\theta$, which further need swap test technique, they are independent of Grover's algorithm. Method 3 is also independent of Grover's algorithm and it works for general unitary operator.

(2). In the above, the reason of only considering the linear combination of real quantum states is that we need the geometrical interpolation of inner product, that is the inner product of $|a\rangle$ and $|b\rangle$ should equal $\cos 2\theta$. However, if we have conjugate operation among quantum states, then we can also study linear combination of complex quantum states  in method 2, 3, 4 by considering the real and imaginary parts respectively. In the following, we only consider the real quantum states.

\section{The searching algorithm based on LCU}
\label{The Grover searching problem}

With the above results, now we can consider the searching problem. Since we only want to explain Grover's algorithm in different perspectives and do not plan to introduce new quantum searching algorithms, we only focus on the searching problem in its simplest case with one marked item, that is $\mathcal{M}=\{x_0\}$ in section \ref{Grover's searching algorithm: brief review}. We should remark that the idea of considering searching problem from LCU was initialized at Long's work \cite{long06}.
Denote
\[\ba{lllll} \vspace{.2cm}
|a\rangle &=& \ds \frac{1}{\sqrt{N}} \sum_{x=0}^{N-1}|x\rangle            &=&\ds \frac{1}{\sqrt{N}} |x_0\rangle+\frac{1}{\sqrt{N}} \sum_{x\neq x_0}|x\rangle, \\
|b\rangle &=& \ds \frac{1}{\sqrt{N}} \sum_{x=0}^{N-1}(-1)^{f(x)}|x\rangle &=&\ds \frac{1}{\sqrt{N}} |x_0\rangle-\frac{1}{\sqrt{N}} \sum_{x\neq x_0}|x\rangle.
\ea\]
They can prepared in time $O(\log N)$.
%Note that the state proportional to $|a\rangle+|b\rangle$ is just $|x_0\rangle$. If $R_{4\theta}^{1/4}$ can be obtained efficiently by above methods, then the searching problem will be solved efficiently, however, this will be a contradiction to the optimality of Grover's algorithm. So generally, if $U$ is efficiently implemented in quantum computer, $U^t$ for $0\leq t\leq 1$ may not efficiently implemented.

In the first method, the norm $\||a\rangle+|b\rangle\|=2/\sqrt{N}$, so the searching problem can be solved in time $O(\sqrt{N}\log N)$.
Actually, this can not be viewed as a method to solve the searching problem, since it applies amplitude amplification technique, which is based on Grover's algorithm.
It is not hard to show that $\cos 2\theta = (2-N)/N$ and $\sin 2\theta = 2\sqrt{N-1}/N$, which means $\theta\approx \pi/2-1/\sqrt{N}$ for large $N$. Although method 2 and 4 rely on swap test, now $\theta$ can be estimated directly.
Based on the analysis of method 2, the Hamiltonian simulation can be efficiently implemented in time $O(T_{\textmd{in}}/\epsilon)$ (see \cite[section 4.7]{nielsen}) where the precision $\epsilon$ is close to $1/\sqrt{N}$. So the searching problem works in time $O(\sqrt{N}\log N)$.
In the third method, the precision $\epsilon$ in quantum phase estimation should choose in size $1/\sqrt{N}$,
so the complexity obtained in the this method is also $O(\sqrt{N}\log N)$. Finally, in the fourth method, to determine $l$ from $4l(\pi/2-1/\sqrt{N}) \mod 2\pi = \pi/2-1/\sqrt{N}$,
we have to solve $(4l-1)/\sqrt{N}=(2m-1/2)\pi$ for some $m$ as minimal as possible (i.e., $m=1$). So $l=O(\sqrt{N})$, which means the complexity in this method is still $O(\sqrt{N}\log N)$. When considering about the query complexity, all the complexities above are just $O(\sqrt{N})$.
%Actually, with a careful calculation on the three methods, we can see that the query complexity is $\lfloor \pi \sqrt{N}/4 \rfloor$, the exact same thing as Grover's algorithm.

These methods also provide us three new understandings about Grover's searching algorithm.
Similar to Grover's algorithm, the above three searching algorithms also apply certain rotations on some fixed initial states. After certain times of rotation, we have a high probability to obtain the marked item. However, searching relies on LCU seems more intuitional, since we know the target states exactly.
The explanation of Grover's algorithm given in \cite{nielsen} based on Hamiltonian simulation aims at finding a suitable Hamiltonian $H$ and a suitable initial state $|\psi\rangle$, such that $e^{-i H t}|\psi\rangle$ is very close to the target $|x_0\rangle$ after $t$ times evolution. The explanations of Grover's algorithm we are given above also share the same essence. Due to the optimality of Grover's algorithm on the searching problem, we can actually say that method 2, 3, 4 are also optimal in solving the implementation of $U^t$ for $0<t<1$. We cannot improve the dependence on the precision anymore.

%With the above results of LCU with two quantum states, now we can consider the Grover searching problem. The notations used below are the sane as section \ref{Grover's searching algorithm: brief review}.
%Denote
%\[\ba{lll} \vspace{.2cm}
%& |a\rangle = \ds \frac{1}{\sqrt{N}} \sum_{x=0}^{N-1}|x\rangle = \ds \frac{1}{\sqrt{N}} \sum_{x\in\mathcal{M}}|x\rangle +\frac{1}{\sqrt{N}} \sum_{x\not\in\mathcal{M}}|x\rangle, \\
%& |b\rangle = \ds \frac{1}{\sqrt{N}} \sum_{x=0}^{N-1}(-1)^{f(x)}|x\rangle = \ds  \frac{1}{\sqrt{N}} \sum_{x\in\mathcal{M}}|x\rangle -\frac{1}{\sqrt{N}} \sum_{x\not\in\mathcal{M}}|x\rangle.
%\ea\]
%They can prepared in time $O(\log N)$.
%Note that the state proportional to $|a\rangle+|b\rangle=\frac{2}{\sqrt{N}} \sum_{x\in\mathcal{M}}|x\rangle$ is equal to $\frac{1}{\sqrt{M}} \sum_{x\in\mathcal{M}}|x\rangle$.
%
%In the first method, the norm $\||a\rangle+|b\rangle\|=2\sqrt{M/N}$, so the searching problem can be solved in time $O(\sqrt{N}\log N)$.
%Actually, this is not a good searching algorithm, since it applies amplitude amplification technique, which is based on Grover's algorithm. So this can not be viewed as a method to solve the searching problem.
%
%It is not hard to show that that $\cos 2\theta = (2M-N)/N$ and $\sin 2\theta = 2\sqrt{M(N-M)}/N$.
%By swap test, we can get an $\epsilon$ approximate of $\theta$ in time $O((\log n)/\epsilon)$.
%The values of $\langle a|k\rangle \langle b|l\rangle-\langle a|l\rangle \langle b|k\rangle$ can be $0,\pm 2/N$.
%And they can be calculated directly.

\section{Linear combination of multiple quantum states}
\label{Linear combination of multiple quantum states}

In this section, we continue the study of the problem considered in section \ref{Four methods to achieve the linear combination of two quantum states}. Now we consider a more general case, that is to achieve the linear combination of multiple quantum states. More precisely, given $m$ positive real numbers $\alpha_j$ and $m$ real quantum states $|x_j\rangle$, which can be prepared efficiently in time $O(T_{\textmd{in}})$, where $j=0,1,\ldots,m-1$, then the problem is how to prepare the quantum state $|y\rangle$ proportional to $y=\sum_{j=0}^{m-1} \alpha_j |x_j\rangle$? And what is the corresponding efficiency? By requiring $\alpha_j$ to be positive does not lose any generality, since we can absorb the negative sign into $|x_j\rangle$.
In the following, we will generalize the four methods studied in section \ref{Four methods to achieve the linear combination of two quantum states} here. We should remark that the LCU technique of Long \cite{long11} is given in a more general form. However, we do not need this general result here. So in the following, we only focus on some simple and straightforward cases.

{\bf Method 1.} The first method given in section \ref{Four methods to achieve the linear combination of two quantum states} still works here at least in two different versions.
The first one \cite{clader} can be viewed as an inspiration of HHL algorithm
\be\ba{lll}\vspace{.2cm} \label{lcu1}
 && \ds\frac{1}{\sqrt{m}}\sum_{j=0}^{m-1}|j\rangle|0,0\rangle\\\vspace{.2cm}
 &\mapsto&\ds\frac{1}{\sqrt{m}}\sum_{j=0}^{m-1}|j\rangle|x_j\rangle|0\rangle \\\vspace{.2cm}
 &\mapsto& \ds\frac{1}{\sqrt{m}}\sum_{j=0}^{m-1}|j\rangle|x_j\rangle\left(t\alpha_j|0\rangle+\sqrt{1-t^2|\alpha_j|^2}|1\rangle\right) \\
 &\mapsto& \ds\frac{t}{m}|0\rangle\sum_{j=0}^{m-1}\alpha_j|x_j\rangle|0\rangle+\textmd{orthogonal parts},
\ea\ee
where $t=1/\max_j|\alpha_j|$.
The first step is the result of control operation to prepare $|x_j\rangle$ with respect to $|j\rangle$; the second step is a control rotation to put $\alpha_j$ into coefficient; the final step is applying Hadamard transformation on the first register.
The success probability equals $\|y\|^2/\max_j|\alpha_j|^2m^2$. So the complexity to get the desired state is
$O(m(T_{\textmd{in}}+\log m)\max_j|\alpha_j|/\|y\|)$.

Another one is given as follows \cite{childs}:
Denote $s=\sum_{j=0}^{m-1}|\alpha_j|$. Define unitary transformation $S$ as $S|0\rangle=\frac{1}{\sqrt{s}}\sum_{j=0}^{m-1}\sqrt{\alpha_j}|j\rangle$.
Then $|y\rangle$ can be obtained from the following procedure:
\be\ba{lcl}\vspace{.2cm} \label{method-1}
|0\rangle|0\rangle &\xrightarrow[]{S\otimes I}& \ds\frac{1}{\sqrt{s}} \sum_{j=0}^{m-1} \sqrt{r_j}|j\rangle|0\rangle \\\vspace{.2cm}
&\rightarrow& \ds\frac{1}{\sqrt{s}} \sum_{j=0}^{m-1} \sqrt{\alpha_j}|j\rangle|x_j\rangle \\
&\xrightarrow[]{S^\dagger\otimes I}& \ds\frac{1}{s} |0\rangle \sum_{j=0}^{m-1} \alpha_j|x_j\rangle +\textmd{orthogonal parts}.
\ea\ee
The second step is also a control operation to prepare $|x_j\rangle$ with respect to $|j\rangle$.
The probability to get $|y\rangle$ equals $\|y\|^2/s^2$, and so the complexity to obtain the desired quantum state is $O(s(T_{\textmd{in}}+C_S)/\|y\|)$, where $O(C_S)$ is the complexity to implement $S$ in quantum computer.

The method (\ref{method-1}) is better than method (\ref{lcu1}), since $s\leq \max_j|\alpha_j|m$ and generally we can believe that $C_S=O(\log m)$. Note that other variants of the above procedure (\ref{method-1}) still exist, however, the above one seems to be the most efficient one. The classical method to this problem also has a similar structure, that is compute $y$ first, then normalize it. Different from the classical one, the quantum method do not need to compute the norm of $y$, since the normalization can be achieved by measurement. More importantly, in the problem of Hamiltonian simulation  \cite{berry,childswiebe}, $y$ is almost a unit vector, so at this time, the complexity is just $O(s(T_{\textmd{in}}+C_S))$. One important special case that we will use in the next section is when the absolute values of the coefficients are close to each other and $\langle x_j|x_k\rangle = \delta_{jk}$, then the complexity now is $O(T+\log m)$.

{\bf Method 2, 3, 4.} As for the other three methods considered in section \ref{Four methods to achieve the linear combination of two quantum states}, the generalization is a little complicate, but still contain a clear geometric explanation. To make things more clear, we first consider the case when $m=2$. Also to make the notations more simple, we rewrite the problem in the form of preparing $|c\rangle$ proportional to
$\alpha |a\rangle +\beta |b\rangle$. Denote the angle between $|a\rangle$ and $|c\rangle$ as $\theta$, the angle between $|a\rangle$ and $|c\rangle$ as $\varphi$. Then we can compute $\theta$ and $\varphi$ from
\[\ba{lll}\vspace{.2cm}
\cos \varphi &=& \langle a|b\rangle, \\
\cos \theta  &=& \ds\frac{\alpha +\beta \langle a|b\rangle}{\|\alpha |a\rangle +\beta |b\rangle\|}.
\ea\]
Now we set $\theta=2t\varphi$ for some $0\leq t\leq 1/2$. Note that at this time, we still have
\[
R_{2\varphi}=(I-2|b\rangle\langle b|)(I-2|a\rangle\langle a|).
\]
So we should calculate $R_{2\varphi}^t=R_\theta$. This reduces to the same problem to implement $U^t$ from $U$. So method 2, 3, 4 can be generalized here directly. The complexity depends on the estimation of $\theta$ and $\varphi$. By swap test, they can be estimated to accuracy $\epsilon$ in time $O(T_{\textmd{in}}/\epsilon)$. The remaining complexity based on method 2, 3, 4 also equals $O(T_{\textmd{in}}/\epsilon)$. Finally, $|c\rangle$ can be obtained in time $O(T_{\textmd{in}}/\epsilon)$ to accuracy $\epsilon$. Here we just use the same precision $\epsilon$ in the two steps of estimation.

As for the general case of preparing $|y\rangle$, we can decompose it into $\log m$ steps. For simplicity, we just set $m=2^k$.
First, we calculate $|\tilde{x}_i\rangle$, which is proportional to $\alpha_{2i}|x_{2i}\rangle+\alpha_{2i+1}|x_{2i+1}\rangle$ in time $O(T_{\textmd{in}}/\epsilon_0)$ to accuracy $\epsilon_0$.
Then we know that $|y\rangle$ is proportional to a new linear summation $\sum_{i=0}^{m/2-1} \tilde{\alpha}_j |\tilde{x}_j\rangle$ for some $\tilde{\alpha}_j$, which can be calculated in time $O(T_{\textmd{in}}/\epsilon_0)$ to accuracy $\epsilon_0$ by swap test.
Continue the above idea, then after $k$ steps, we can get $|y\rangle$. Note that, in each step, the input complexity to obtain the new linear combination updates by a factor $1/\epsilon_0$. By induction, $|y\rangle$ can be obtained in time
\[
\sum_{i=0}^{k-1} {2^i T_{\textmd{in}}}/{\epsilon_0^{k-i}} \approx {T_\textmd{in}}/{\epsilon_0^k} = {T_\textmd{in}}/{\epsilon_0^{\log m}}
= T_\textmd{in}m^{\log 1/\epsilon_0}.
\]
In the above equality analysis, we have assumed that $\epsilon_0<1/2$.
After $k$ steps, the error is enlarged into $m\epsilon_0$. So to make the final error is small in size $\epsilon$, we should choose $m\epsilon_0=\epsilon$. Finally we know that the complexity to prepare $|y\rangle$ is
$O(T_\textmd{in}m^{\log m/\epsilon})$. In the following, we list all the LCU methods introduced above for comparison of their efficiency and application scopes.

{\renewcommand\arraystretch{1.7}
\begin{table}[htb]
\centering
\caption{
Comparison of different methods to solve the linear combination of $m$ quantum states $|x_j\rangle$, where $O(T_{\textmd{in}})$ in the complexity to prepare $|x_j\rangle$, $\{\alpha_j\}_{0\leq j\leq m-1}$ are real numbers, $y=\sum_j \alpha_j |x_j\rangle$, $s=\sum_j |\alpha_j|$ and $O(C_S)$ is the required time to implement the unitary $S$ with $S|0\rangle=\frac{1}{\sqrt{s}}\sum_{j=0}^{m-1}\sqrt{|\alpha_j|}|j\rangle$.
}
\begin{tabular}{cc}\hline\hline
   \hspace{.9cm}{\bf Method} \hspace{.9cm} & \hspace{1.5cm} {\bf Complexity}  \hspace{1.4cm} \\   \hline
  Method 1 (version 1)  & $O(m(T_{\textmd{in}}+\log m)\max_j|\alpha_j|/\|y\|)$ \\
  Method 1 (version 2)  & $O(s(T_{\textmd{in}}+C_S)/\|y\|)$ \\
  Method 2, 3, 4        & $O(T_\textmd{in}m^{\log (m/\epsilon)})$ \\ \hline\hline
\end{tabular}
\label{table1}
\end{table}}

As we can see from table \ref{table1}, although the complexity in method 2, 3, 4 contains a factor $m^{\log (m/\epsilon)}$, it is independent of $\alpha_j$ and $\|y\|$.
When $m$ is large, such as in the Hamiltonian simulation problem \cite{berry,childswiebe} or in the quantum linear system algorithm \cite{childs-linear-system} that achieves exponential speedup on precision, then method 1 will be better than the other three.
However, if $m$ is small, such as in the quantum iteration algorithms considered in \cite{kerenidis,rebentros,shao},
then method 2, 3, 4 can play more important roles than method 1. Method 2, 3, 4 also work better when $\max_j|\alpha_j|$ is very large, such as the Arnoldi method considered in \cite{shao} and the quantum state preparation problem considered in the next section.

\section{Application of LCU in quantum state preparation}
\label{Application of LCU in quantum state preparation}

Let $x=(x_0,\ldots,x_{n-1})$ be a real vector, the quantum state it corresponds to equals $|x\rangle=\frac{1}{\|x\|} \sum_{i=0}^{n-1} x_i|i\rangle$. The transformation from classical data $x$ into its quantum state $|x\rangle$ is usually called the ``input problem" in quantum computer \cite{biamonte}, which forms the initial step in many quantum algorithms, such as \cite{childs-linear-system,clader,harrow,kerenidis,kerenidis-iteration,lloyd,rebentros,rebentros-newton,rebentros-svd,wang,wiebe,wossnig}.  In the following, we first show two simple quantum algorithms to solve this problem. Then as an inspiration of the second algorithm, we propose two new quantum algorithms based on LCU with high efficiency.

The most naive method is defining a unitary $U$ such that $U|0\rangle=|x\rangle$. The efficiency of preparing $|x\rangle$ is totally determined by $U$. In the worst case, $U$ can be implemented in time $O(n^2(\log n)^2\log^c(n^2(\log n)^2/\epsilon))$ to precision $\epsilon$ in quantum computer, where $c$ is some constant close to 2  (see \cite[Chapter 4]{nielsen}). So we can prepare $|x\rangle$ within the same time in the worst case. Conclude this, we have

\bp \label{prop1}
For any vector $x$, its quantum state can be prepared in time $O(n^2(\log n)^2\log^c(n^2(\log n)^2/\epsilon))$ to precision $\epsilon$ in quantum computer.
\ep

Although the above method works for all cases, it is not efficient generally. And we still hope there exist more efficient quantum algorithms to solve the input problem even in some special cases. Under certain conditions, the input problem can actually solved efficiently in polynomial time, for instance, see \cite{clader,grover-state-preparation,lloyd,soklakov}. In the following, we focus on one of them, which can be viewed as an application of method (\ref{lcu1}) proposed in \cite{clader}. Note that in preparing the quantum state of $x$, we actually only need to focus on its nonzero entries, so in the following, we assume that all entries of $x$ are nonzero. However, the results obtained below hold for all vectors. To prepare the quantum state of $x$ based on (\ref{lcu1}), we just need to choose $|x_j\rangle = |j\rangle$. It is easy to obtain the following result from the complexity of  (\ref{lcu1})

\bp \label{prop2}
For any vector $x=(x_0,\ldots,x_{n-1})$, its quantum state can be prepared in time $O(\kappa(x)\log n)$,
where $\kappa(x)={\max_k |x_k|}/{\min_{k,x_k\neq 0}|x_k|}$.
\ep

As we can see from the above result, the quantum state preparation algorithm works efficient when $x$ is a relatively uniform distributed vector, that is $\kappa(x)=O({\rm poly}(\log n))$. One way to grasp this property is decomposing $x$ into a linear summation of several relatively uniform distributed vectors. In the following, we give two different such decompositions.

Let $x=(x_0,\ldots,x_{n-1})$ be a real vector. For simplicity, we assume that $|x_0|=\min_{k,x_k\neq 0}|x_k|$. Find the minimal $q$ such that $\kappa(x)\leq 2^q$, so $q\approx \log \kappa(x)$. For any $1\leq j\leq q$, there are several entries of $x$ such that their absolute values lie in the  interval $I_j=[2^{j-1}|x_0|,2^{j}|x_0|)$. Define $y_j$ as the $n$ dimensional vector by filling these entries into the corresponding positions of them in $x$ and zero into other positions. Then $x=y_1+\cdots+y_q$. For any $j$, we have $\kappa(y_j)\leq 2$, so the quantum state $|y_j\rangle$ of vector $y_j$ can be prepared efficiently in time $O(\log n)$ by proposition 2. We also have
$|x\rangle=\lambda_1|y_1\rangle+\cdots+\lambda_q|y_q\rangle$, where $\lambda_j=\|y_j\|/\|x\|$. From the method (\ref{method-1}) given in the above section, the complexity to achieve such a linear combination to get $|x\rangle$ equals
\[\ba{lll} \vspace{.2cm}
\ds O\Bigg((\log n)\sum_{j=1}^q \frac{\|y_j\|}{\|x\|}\Bigg)
&=& \ds O\left(\sqrt{q}(\log n)\right) \\
&=& O(\sqrt{\log\kappa(x)}(\log n)),
\ea\]
where the first identity is because of the relation between 1-norm and 2-norm of vectors, more precisely, it is a result of
$\sum_{j=1}^q \|y_j\| \leq \sqrt{q} \sqrt{\sum_{j=1}^q \|y_j\|^2}=\sqrt{q}\|x\|$. Therefore, we have

\bt \label{thm1}
Let $x=(x_0,\ldots,x_{n-1})$ be a given vector and $\kappa(x)=\max_k |x_k|/\min_{k,x_k\neq 0} |x_k|$. Then its quantum state can be prepared in time $O(\sqrt{\log\kappa(x)}(\log n))$.
\et

This result achieves exponential speedup than the algorithm given in proposition \ref{prop2}. If $\kappa(x)$ is too large and small error is allowed in preparing $|x\rangle$, then we may consider giving up the entries that are close to $\min_{k,x_k\neq 0} |x_k|$ if there are not too many of them. So  $\kappa(x)$ is large in a reasonable sense. Moreover, even if $\max|x_k|=2^{1000}$ and $\min_{k,x_k\neq 0} |x_k|=1$ for instance, then $\log \kappa(x)=1000$, which is still an reasonable small constant. From these points, the above result seems to be a pretty good algorithm to prepare quantum states.

Another decomposition is more direct and easy. In this case, the new LCU technique we proposed will play a central role. The corresponding result is much better than theorem \ref{thm1}. Assume that all entries of $x$ are nonzero.
Define
$y=M(\textmd{sign}(x_0),\ldots,\textmd{sign}(x_{n-1})),$
where $M\geq\max_k|x_k|$ and $\textmd{sign}(x_i)=1$ if $x_i>0$; $\textmd{sign}(x_i)=-1$ if $x_i < 0$.
Then the quantum state $|y\rangle$ of $y$ can be obtained efficiently in time $O(\log n)$. Also define
$z:=x+y=(\textmd{sign}(x_0)M+x_0,\ldots,\textmd{sign}(x_{n-1})M+x_{n-1}),$
which is uniformly distributed with $n$ nonzero entries. By proposition \ref{prop2}, the quantum state $|z\rangle$ of $z$ can be obtained efficiently in time $O(\log n)$ too. So we obtain the following decomposition about $|x\rangle$:
\[
|x\rangle = \frac{1}{\|x\|} (z-y) = \frac{\|z\|}{\|x\|}|z\rangle - \frac{\|y\|}{\|x\|}|y\rangle.
\]

What we should do next is achieving the linear combination of two efficiently prepared quantum states by LCU.
By method 1 given in section \ref{Four methods to achieve the linear combination of two quantum states},
$|x\rangle$ can be prepared in time
$
O(\sqrt{{\|y\|^2+\|z\|^2}/{\|x\|^2}}(\log n)).
$
This is pretty large. More precisely
\[
\kappa(z)=\frac{\max_k (|\textmd{sign}(x_k)M+x_k|)}{\min_k(|\textmd{sign}(x_k)M+x_k|)}=\frac{M+\max_k|x_k|}{M+\min_k|x_k|}.
\]
To make $\kappa(z)$ a small constant, we can just choose $M=\max_k|x_k|$. Note that $\|x\|^2\geq n\min_k|x_k|^2$, so
\[
\frac{\|y\|^2+\|z\|^2}{\|x\|^2}
=\frac{2nM^2+2M\sum_i x_i+\|x\|_2^2}{\|x\|_2^2}
\leq 3\kappa(x)^2+1.
\]
Hence, the complexity obtained by this decomposition is $O(\kappa(x)(\log n))$, which is the same as proposition \ref{prop2}.
The new LCU technique proposed in the section \ref{Four methods to achieve the linear combination of two quantum states} is independent of the influence of $\lambda$ and $\mu$. So we can get $|x\rangle$ in time $O((\log n)/\epsilon)$, where $\epsilon$ is the precision. 
Conclude this, we have

\bt \label{thm2}
For any vector $x$, its quantum state can be prepared in time $O((\log n)/\epsilon)$ to precision $\epsilon$.
\et

%The following table summarizes the four quantum algorithms to achieve quantum state preparation in the general case. Note that the algorithms given in proposition \ref{prop2} and theorem \ref{thm1} contain no errors in the quantum state of $x$.

%{\renewcommand\arraystretch{1.7}
%\begin{table}[htb]
%\centering
%\caption{
%Comparison of different quantum algorithms to prepare quantum state of $x=(x_0,\ldots,x_{n-1})$,
%where  $\kappa(x)=\max_k |x_k|/\min_{k,x_k\neq 0} |x_k|$ and $\epsilon$ is the precision, $c\approx 2$.
%}
%\begin{tabular}{cc}\hline\hline
%   \hspace{1.4cm}{\bf Algorithms given in} \hspace{1.4cm} & \hspace{3cm} {\bf Complexity}  \hspace{3cm} \\   \hline
%   Proposition \ref{prop1}    & $O(n^2(\log n)^2\log^c(n^2(\log n)^2/\epsilon))$ \\
%   Proposition \ref{prop2} & $O(\kappa(x){(\log n) })$ \\
%   Theorem \ref{thm1} & $O(\sqrt{\log \kappa(x)}(\log n)) $ \\
%   Theorem \ref{thm2} &  $O((\log n)/\epsilon^{1.6})$ \\ \hline\hline
%\end{tabular}
%\label{table2}
%\end{table}}

\section{Conclusion}

In this paper, we proposed a new quantum algorithm to achieve LCU problem. It performs better than the old LCU when the number of unitaries is small or the linear coefficients are very large. As an application, we showed that the quantum state of any real vector can be prepared efficiently in polynomial time in quantum computer. Also three new explanations about Grover's algorithm are obtained based on the new LCU technique.

However, the efficiency of the new LCU technique still needs to improve, since it is not polynomial in the number of unitaries. Moreover, it is worth to find more applications of LCU, old or new. One possible researching direction is Krylov iteration methods. Most Krylov iteration methods, such as Lanczos, CG, Arnoldi are very efficient, so the iteration step is not large generally. Also Krylov iteration methods are important modern iteration methods to solve large linear systems $Ax=b$ and to estimate eigenvalues of large matrices. The basic idea is approximating $A^{-1}$ by a polynomial of $A$. So HHL algorithm or SVE and LCU techniques will play important roles in generalizing classical Krylov iteration methods into quantum versions. The efficiency can achieve exponential speedup at least about the dimension of $A$ based on HHL algorithm or SVE. If LCU is efficient, then we also make the complexity of quantum Krylov iteration methods depend on the iteration steps in polynomial form. Finally, the efficiency of the classical Krylov iteration methods will be improved greatly in their quantum versions.

%\section{Acknowledgement}
{\bf Acknowledgement.}
This work is supported by the NSFC Project 11671388 and the CAS Frontier Key Project QYZDJ-SSW-SYS022.

\end{document}